\documentclass[twocolumn,showpacs,preprintnumbers,amsmath,amssymb,superscriptaddress]{revtex4}

\usepackage{graphicx}
\usepackage{bm}

\newcommand\Tm{\overline{\mathbf{T}}}
\newcommand\rr{{\bf r}}

\begin{document}
\title{Optimizing persistent random searches}

\author{Vincent Tejedor}
\affiliation{Physics Department, Technical University of Munich, James Franck
Strasse, 85747 Garching, Germany}
\affiliation{Laboratoire de Physique Th\'eorique de la Mati\`ere Condens\'ee, CNRS/Universit\'e Pierre et Marie Curie, 4 Place Jussieu, 75005
Paris}

\author{Raphael Voituriez}
\affiliation{Laboratoire de Physique Th\'eorique de la Mati\`ere Condens\'ee, CNRS/Universit\'e Pierre et Marie Curie, 4 Place Jussieu, 75005
Paris}

\author{Olivier B\'enichou}
\affiliation{Laboratoire de Physique Th\'eorique de la Mati\`ere Condens\'ee, CNRS/Universit\'e Pierre et Marie Curie, 4 Place Jussieu, 75005
Paris}


\begin{abstract}
We consider a minimal model of persistent random searcher  with short range memory. 
We calculate exactly for such searcher the mean first-passage time  to a target in a bounded domain  and find that it admits a non trivial minimum as function of the persistence length.  This reveals an optimal search strategy which differs markedly 
from the simple ballistic motion obtained in  the case of Poisson distributed targets.  Our results  show that the distribution of targets plays a crucial role in the random search problem.  In particular, in the biologically relevant cases  of either a single target or regular   patterns of targets, we find that, in strong contrast with  repeated statements in the literature,  persistent random walks with exponential distribution of excursion lengths can minimize the search time,  and in that sense perform better than any Levy walk.

\end{abstract}


\maketitle


The random search problem addresses the question of determining the time it takes a searcher performing a random walk to find a target \cite{Benichou:2011fk}.
At the microscopic scale, search processes naturally occur in the context of chemical reactions, for which the encounter of reactive molecules is a required first step. An obvious historical example is the theory of diffusion---controlled reactions, which has regained interest in the last few years in the context of  genomic transcription in cells \cite{Benichou:2011fk}.  Interestingly, the random search problem has also proved in the last decades to be relevant at the macroscopic scale, as in the case of animals searching for a mate, food, or shelter \cite{Bell:1991,Klafter:1996, Viswanathan:1996,Viswanathan:1999a,Benichou:2005qd,Oshanin:2007a,Friedrich:2008kx,Lomholt:2008}.

In all these examples, the time needed to discover a target is a limiting quantity, and consequently the minimization of this search time often appears as essential. In this context Levy walks, which are defined as randomly reoriented ballistic excursions whose length $l$ is drawn from a power law distribution $P(l)\underset{l \to \infty}{\propto} 1/ l^{1+\mu} $ with $0 < \mu \leq 2$,  have been suggested as potential candidates of optimal strategies \cite{Viswanathan:1999a}. In fact, Levy walks  have been shown mostly numerically to optimize the search efficiency, but only in the particular case where the targets are distributed in space according to a Poisson law, and are in addition  assumed to  regenerate at the same location after a finite time. Conversely, in the case of a destructive search where each target can be found only once the optimal strategy proposed in \cite{Viswanathan:1999a} is not anymore of Levy type, but reduces to a trivial ballistic motion. Given these restrictive conditions of optimization, the potential selection by evolution of Levy trajectories as optimal search strategies is  disputable, and in fact the field observation of Levy  trajectories for foraging animals  is still elusive  and controversial \cite{Edwards:2007,Benhamou:2007fk,James:2011uq}.

From the theoretical point of view, the search time can be quantified as the first-passage time of the random searcher to the target \cite{Redner:2001a}. In the case of a single target in a bounded domain, or equivalently of infinitely many regularly spaced targets, asymptotic  results  for the mean first-passage time (MFPT) and the full distribution of the first-passage time have been obtained for Markovian scale invariant random walks \cite{Condamin:2007zl, BenichouO.:2010}. These results  apply in particular directly to Brownian particles that are subject to thermal fluctuations,  and therefore to diffusion--limited reactions in general. At larger scales however, most examples of searchers -- even if random --  have at least short range memory skills and show persistent motions, as is the case for bacteria \cite{bergcoli} or larger organisms \cite{Bell:1991}, which cannot be described as Markovian scale invariant processes. The study of  persistent random walks  therefore appears as  crucial  to assess the  efficiency of many search processes, and has actually also proved to be important in various fields such as neutron or light scattering \cite{Blanco:2003a,Mazzolo:2004a,Zoia:2011vn}. In this context exact results have been derived  that characterize the diffusion properties of persistent walks in infinite space  \cite{Ernst:1988fk,Gilbert:2011ys,Zoia:2011vn}, or  mean return times in bounded domains \cite{Blanco:2003a,Mazzolo:2004a,Benichou:2005a,Condamin:2005qr}. The question of determining first-passage properties of persistent walks has however remained  unanswered so far. 
\begin{figure}[htb!]
\includegraphics[width = 0.6\linewidth,clip]{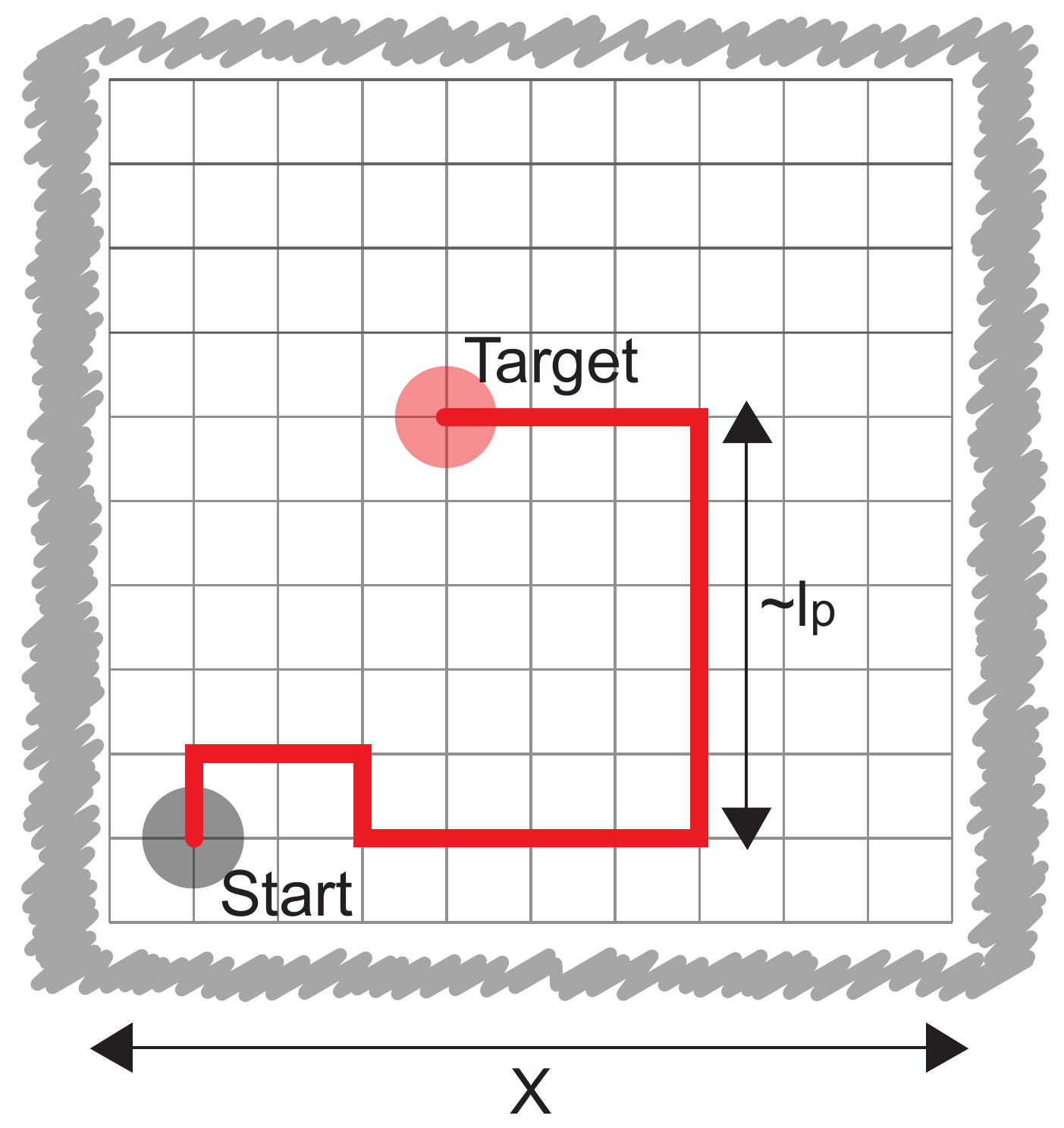} 
\caption{(Color online) Example of search trajectory for a persistent random searcher in a bounded domain.}
\label{dessin}
\end{figure}

In this paper, we consider a minimal model of persistent random searcher -- called persistent random walk model hereafter -- with short range memory  characterized by an exponential distribution of the length of its successive ballistic excursions $P(l)\underset{l \to \infty}{\propto} e^{-\alpha l/l_p} $, where $l_p$ is the persistence length of the walk and $\alpha$ a numerical factor. We calculate exactly for such persistent random walker the MFPT to a target in a bounded domain, which corresponds implicitly to the case of a destructive search since the target can be discovered only once,  and find that it admits a non trivial minimum as a function of $l_p$, thus revealing an optimal search strategy which 
is very different from  the simple ballistic motion obtained in  the case of Poisson distributed targets.  In addition, we show numerically that such  optimal persistent random walk strategy is more efficient than {\it any } Levy walk of parameter $\mu\in ]0,2[$. Together, our results  show that the distribution of targets plays a crucial role in the random search problem.  In particular, in the biologically relevant cases  of either a single target or   patterns of targets characterized by a peaked distribution of the target to target distance \cite{Bell:1991}, we find that, in marked  contrast with  repeated statements in the literature,  persistent random walks with exponential distribution of excursion lengths can minimize the search time,  and in that sense perform better than any Levy walk.


The model is defined as follows (see Fig. (\ref{dessin})).  We  consider a persistent random walker in discrete time and space, moving on a $d$--dimensional cubic lattice $\mathcal{L}$ of volume $V=X^d$, where a single target site is located. In practice we  take $d=2$ or $d=3$, make use of periodic boundary conditions   and consider the dilute regime $X\gg 1$. This geometry  encompasses both  cases of a single target centered in a confined domain, and  of regularly spaced targets in infinite space with concentration $1/V$. The latter situation can be seen as a limiting case of target distribution with strong correlations, as opposed to the Poissonian case, and  is biologically meaningful for example in the case of repulsive interactions between targets \cite{Bell:1991}. The case of a target of arbitrary position in a domain with reflective boundary conditions can also be solved exactly using similar techniques and has been checked to yield analogous results; analytical expressions are however much more complicated in this case and are omitted here for clarity.  Note that here the lattice step size corresponds to the target size and is set to 1, which defines the unit length of the problem.  At each time  step, the random searcher has a probability $p_1$ to continue in the same direction, $p_2$ to go backward, and $p_3$ to choose an orthogonal direction, so that $p_3 = (1-p_1-p_2)/(2 d-2)$. Following \cite{Ernst:1988fk}, we  denote $p_1 = p_3 + \epsilon$ and $p_2 = p_3 - \delta$, and set in what follows $\delta=0$ for the sake of simplicity. The probability of a ballistic excursion of $l$ consecutive steps with unchanged direction is then $P(l)=(1-p_1)p_1^{l-1}$, and the persistence length of the walk can be defined as $l_p=\sum_{l=1}^{\infty}l P(l)=1/(1-p_1)$ where $p_1=(1+(2d-1)\epsilon)/(2d)$, so that eventually  $l_p=(2d/(2d-1))/(1-\epsilon)$. In what follows we calculate analytically the search time $\langle \Tm \rangle$, defined here as the MFPT to the target averaged over all possible starting positions and velocities of the searcher, and analyze   its dependence on the persistence length $l_p$ (or equivalently $\epsilon $) and the volume $X^d$.

While the position process alone  is non Markovian, the joint process of the position and velocity of the searcher is Markovian. One can therefore derive an exact backward equation for the MFPT  $\Tm(\rr,{\bf e}_i)$ to the target of position $\rr_T$, for a random searcher starting from $\rr$ with initial velocity ${\bf e}_i$, where $ \mathcal{B}=\{{\bf e}_1,\ldots,{\bf e}_d\}$ defines a basis of the lattice :
\begin{eqnarray}
\Tm(\rr,{\bf e}_i) =  p_1 \Tm(\rr+{\bf e}_i,{\bf e}_i) + p_2 \Tm(\rr-{\bf e}_i,-{\bf e}_i) \nonumber \\
 + p_3\!\!\!\!\!\! \sum_{{\bf e}_j \in \mathcal{B},j \neq i}\!\!\!\!\!\! \left ( \Tm(\rr+{\bf e}_j,{\bf e}_j) + \Tm(\rr-{\bf e}_j,-{\bf e}_j) \right ) + 1. \ \label{eq:Kolmo2}
\end{eqnarray}
Note that this equation holds for all sites $\rr\not=\rr_T$. Indeed, by definition for $\rr=\rr_T$ the lhs of of Eq. (\ref{eq:Kolmo2}) yields $\Tm(\rr_T,{\bf e}_i)=0$, while the rhs  gives  the mean return time to site $\rr_T$, which is exactly equal to $V$ in virtue of a theorem due to Kac \cite{Aldous:1999}. We next introduce the Fourier transform $\widetilde{f}({\bf q})= \sum_{\rr\in\mathcal{L}} f({\rr})e^{-\imath {\bf q}.\rr}$ of  a function $f(\rr)$, where $q_i =2\pi n_i/X$ with $n_i\in [0,X-1]$.  The  Fourier transform of Eq.(\ref{eq:Kolmo2}), completed by the above discussed result at $\rr=\rr_T$ then yields :
\begin{equation}
\widetilde{\Tm}({\bf q},{\bf e}_i) + V e^{- \imath {\bf q}.\rr_T} = \epsilon \widetilde{\Tm}({\bf q},{\bf e}_i) e^{\imath {\bf q}.{\bf e}_i}  + V \delta({\bf q}) + p_3g({\bf q})
\end{equation}
where 
\begin{equation}
g({\bf q}) = \sum_{{\bf e}_j \in \mathcal{B}} \left ( \widetilde{\Tm}({\bf q},{\bf e}_j) e^{ \imath  {\bf q}.{\bf e}_j}  + \widetilde{\Tm}({\bf q},-{\bf e}_j) e^{- \imath  {\bf q}.{\bf e}_j}  \right ),
\end{equation}
and $\delta({\bf q})$ is the $d$-dimensional Kronecker function. We thus obtain:
\begin{equation}\label{Tfourier}
\widetilde{\Tm}({\bf q},{\bf e}_i)=\frac{V \left ( \delta({\bf q}) - e^{- \imath  {\bf q}.\rr_T} \right ) + p_3 g({\bf q})}{1- \epsilon e^{ \imath  {\bf q}.{\bf e}_i}}.
  \end{equation}
Summing Eq.(\ref{Tfourier}) times $e^{ \imath  {\bf q}.{\bf e}_i}$ over all ${\bf e}_i$ yields a closed equation for $g({\bf q})$, which is solved by:
\begin{equation}
g({\bf q}) = \frac{\gamma({\bf q},\epsilon)V \left ( \delta({\bf q}) - e^{- \imath  {\bf q}.\rr_T} \right )}{1-p_3\gamma({\bf q},\epsilon)}
\end{equation}
where
\begin{equation}
\gamma({\bf q},\epsilon)= 2\sum_{{\bf e}_j \in \mathcal{B}} \frac{\cos ( {\bf q}.{\bf e}_j )-\epsilon}{1 + \epsilon^2 - 2 \epsilon \cos (  {\bf q}.{\bf e}_j)}.
\end{equation}
Substituting this expression of $g({\bf q})$ in Eq.(\ref{Tfourier})
 then   leads to an  explicit expression of $\widetilde{\Tm}({\bf q},{\bf e}_i)$. After Fourier inversion and  averaging over all possible starting positions and velocities we finally obtain after some algebra:
\begin{equation}\label{exactT}
\langle \Tm \rangle = \frac{-\epsilon(V-1)}{1-\epsilon }  + \frac{1+\epsilon^2}{1-\epsilon^2} \sum_{{\bf q}\neq {\bf 0}}\frac{1}{1-h({\bf q},\epsilon)} 
\end{equation}
where
\begin{equation}
h({\bf q},\epsilon)= \frac{(\epsilon-1)^2}{d}\sum_{{\bf e}_j \in \mathcal{B}} \frac{\cos ( {\bf q}.{\bf e}_j )}{1 + \epsilon^2 - 2 \epsilon \cos (  {\bf q}.{\bf e}_j)}
\end{equation}
and $\sum_{{\bf q}\neq {\bf 0}}$ denotes the sum over all possible vectors ${\bf q}$ defined above except ${\bf q}={\bf 0}$. This exact expression of the search time for a non Markovian searcher  constitutes the central result of this paper.  We discuss below its physical implications, based on two useful approximations.

\begin{figure}[htb!]
\includegraphics[width = 0.9\linewidth,clip]{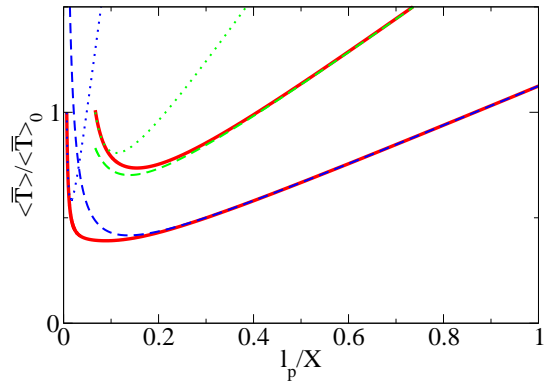} 
\caption{(Color online) Search time  for a 2--dimensional persistent random walker $\langle \Tm \rangle$ normalized by the search time for a non persistent walker ($\langle \Tm \rangle_0$) as a function of the rescaled persistence length, for $X=10$ (upper set of curves) and $X=100$ (lower set of curves). The red line stands for the exact result of Eq. (\ref{exactT}), the dotted lines for the approximation $\epsilon \ll 1$ of Eq. (\ref{smallepsilon}), the dashed lines for the approximation $\epsilon \to 1$ of Eq. (\ref{eq:MFPT2}). We used the identity $l_p=(2d/(2d-1))/(1-\epsilon)$.}
\label{fig:eps}
\end{figure}

We first consider the case where $\epsilon\ll 1$, which implies that the persistence length is of the same order as the target size ($ l_p=\mathcal{O}(1)$).  In this regime the search time reads:
\begin{equation}
\langle \Tm \rangle \! \underset{\epsilon \ll 1}{=} \!\! A(\epsilon,V) (V-1) + \frac{1}{D(\epsilon)} \langle \Tm \rangle_0, \label{smallepsilon}
\end{equation}
where $\langle \Tm \rangle_0$ is the search time of a non persistent random walk ($\epsilon=0$) which is known exactly \cite{Condamin:2006fp}. The quantity $A(\epsilon,V)$ writes 
$
A(\epsilon,V) =(B_d(V)-1) \epsilon+\mathcal{O}(\epsilon^2)
$ 
where $B_d (V)$ depends on $V$ as follows: 
\begin{equation}
B_d(V) =  \frac{2}{V} \sum_{{\bf q}\neq {\bf 0}} \frac{\displaystyle \frac{1}{d} \sum_{{\bf e}_j \in \mathcal{B}} \left (1-\cos ( 2 \pi {\bf q}.{\bf e}_j ) \right )^2}{\displaystyle \left ( \frac{1}{d} \sum_{{\bf e}_j \in \mathcal{B}} 1-\cos ( 2 \pi {\bf q}.{\bf e}_j ) \right )^2}.
\end{equation}
 In the  dilute regime ($V \to \infty$), $B_d$ has a finite limit (for example  $B_2 \simeq 2.72$) and Eq.(\ref{smallepsilon}) provides a useful approximate of the search time.  In this expression,   $D(\epsilon) = (1+\epsilon)/(1-\epsilon)$ is  the  diffusion coefficient of the persistent random walk normalized by the diffusion coefficient of the non persistent walk (case $\epsilon=0$)  \cite{Ernst:1988fk}. Hence, in Eq. (\ref{eq:MFPT2}), $ \langle \Tm \rangle_0/D$ is the search time expected for a non persistent random searcher of same normalized diffusion coefficient $D$. Note that the persistence property yields a non trivial additive correction which scales linearly with the volume, and therefore should not be neglected;  this could be related to the "residual" mean first passage time described in \cite{Tejedor:2011uq}.  
As shown in Fig. \ref{fig:eps}, the approximation of Eq. (\ref{smallepsilon}) is  accurate as long as $l_p$ is small (that is $\epsilon\ll 1$).

We next consider the case where the persistence length is much larger than the target size, that is $ l_p\gg 1$, or equivalently $\epsilon\to 1$.  In this regime the search time reads in the case $d=2$ :
\begin{eqnarray}
 \label{eq:MFPT2}
\langle \Tm \rangle &  = & \frac{2(X-1)}{1-\epsilon} +\frac{(X-1)^2}{2} \\ \nonumber
  &    & +(1-\epsilon) \frac{(X-1)(X+3)(X-2)}{12} +
\mathcal{O}\left ( (1-\epsilon)^2 \right ).
\end{eqnarray}
Fig. \ref{fig:eps} shows that this expression provides a good approximation  of  the exact result of Eq. (\ref{exactT}) for $l_p\gg 1$. Note that the search time diverges for $l_p\to\infty$ (or  $\epsilon\to 1$) because the searcher can then be trapped in extremely long unsuccessful ballistic excursions.
\begin{figure}[htb!]
\includegraphics[width = 0.9\linewidth,clip]{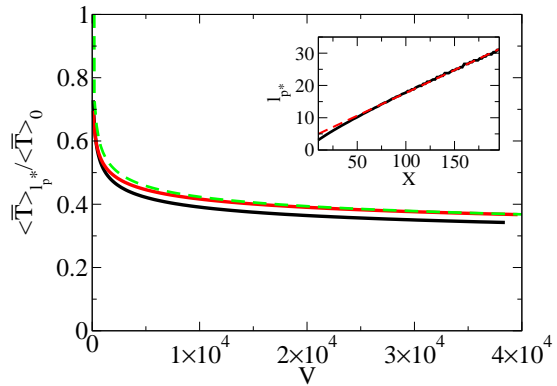} 
\caption{(Color online) Optimal search time scaled by the non persistent case as the function of the domain volume $V$ for $d=2$. The black line stands for the numerical optimization of Eq. (\ref{exactT}), the red line for the analytical optimization of Eq. (\ref{eq:MFPT2}), and the green dashed line for a fit  $A/\ln(V)$, where we used the identity $l_p=(2d/(2d-1))/(1-\epsilon)$. Inset: The black line stands for the persistence length at the minimum, $l_p^*$, obtained by a numerical optimization of Eq. (\ref{exactT}), as a function of $X$. The dashed red line is a linear fit of this curve ($l_p \simeq 0.14 \  X + 3.6$).}
\label{fig:gain-2D}
\end{figure}

Both asymptotics $\epsilon\to 0$ and $\epsilon\to 1$ clearly show that the search time can be minimized as a function of $\epsilon$ or equivalently $l_p$, as seen in  Fig. \ref{fig:eps}. The minimum can be obtained from the analysis of the exact expression (\ref{exactT}), and reveals that the search time is minimized in the case $d=2$ for $l_p=l_p^*\underset{X \to \infty}{\sim} \lambda_2  X$ with $\lambda_2\simeq0.14...$. Note that the asymptotic expression (\ref{eq:MFPT2}) yields a good analytical approximate of this minimum.  This defines the optimal strategy for a persistent random searcher, which is realized when the persistence length has the same order of magnitude as the typical system size. In particular, for large system sizes the optimal persistence length becomes much larger than the target size. We stress however that the numerical factor $\lambda$ is non trivial and notably small. This optimal strategy can be understood as follows. In the regime $l_p\ll X$, the random walk behaves as a regular diffusion and is therefore  recurrent for $d=2$. The exploration of space is therefore redundant and yields a  search time that scales in this regime  as $V\ln V$ \cite{Tejedor:2009}. On the contrary for $l_p\gg 1$ exploration is transient at the scale of $l_p$ and therefore less redundant. As soon as $l_p\sim X$ one therefore expects the search time to scale as $V$ \cite{Tejedor:2009}. Taking $l_p$ too large  however becomes unfavorable since the searcher can be trapped in extremely long unsuccessful ballistic excursions, so that one indeed expects an optimum in the regime  $l_p\sim X$. This  argument suggests the following scaling of the  optimal search time scaled by the non persistent case in the case $d=2$ :  
\begin{equation}
 \frac{\langle \Tm\rangle_{l_p^*}}{\langle \Tm\rangle_ {0}} \propto 1/\ln(V),
\end{equation}
which can indeed be derived from the asymptotic expression (\ref{eq:MFPT2}) (see  also Fig. \ref{fig:gain-2D}). This shows the efficiency of the optimal persistent search strategy in the large volume limit, as compared to the non persistent Brownian strategy. 
Note that for $d =3$ a similar analysis applies. In particular the search time is minimized for a value of the persistence length that again grows linearly with the system size $l_p^*\underset{X \to \infty}{\sim} \lambda_{3}  X$ with however a slightly different numerical value of the coefficient $\lambda_{3}\simeq0.12...$. Additionally, since for $d =3$ one has $\langle \Tm\rangle_{0}\propto V$,   the  scaled optimal search time then tends to a constant.

\begin{figure}[htb!]
\includegraphics[width = 0.9\linewidth,clip]{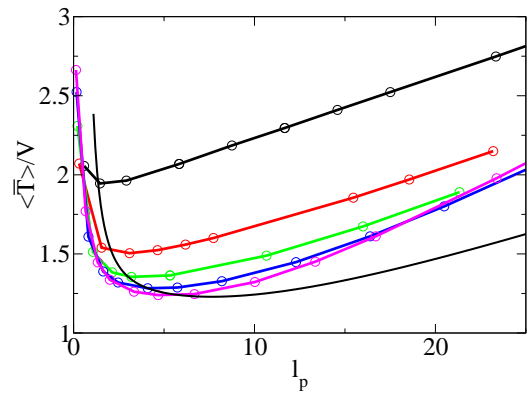} 
\caption{(Color online) Numerical computation of the search time  for a Levy walk  on a  2D lattice ($X=50$). Plots with circles stand for the search time for the following values of  $\mu$ (from top to bottom): $\mu = 1.2$, 1.4, 1.6, 1.8 and 2. The black line stands for a persistent random walk for several values of $\epsilon$. The abscissa   stands for the persistence length $l_p$, function of $c$ or $\epsilon$.}
\label{fig:levy}
\end{figure}

Last, we compare the efficiency of the persistent and Levy walk strategies. More precisely we consider a Levy walker such that the  distribution of the length of its successive ballistic excursions  follows a  symmetric  Levy law of index $\mu$ and scale parameter $c$ restricted to the positive axis, defined by the Fourrier transform $\widehat{P}(k)=e^{-c|k|^\mu}$, so that $P(l)\underset{l \to \infty}{\propto} 1/ l^{1+\mu} $. For $0<\mu\le 1$, the persistence length is infinite, yielding in turn an infinite    search time. We therefore focus on the regime  $1<\mu\le 2$ and study numerically the dependence of the search time on both $\mu$ and $l_p$ (which is set by $c$). Figure \ref{fig:levy} shows that the search time can be minimized as a function of $l_p$ for all $\mu\in ]1,2]$, and that this optimal value decreases when $\mu$ is increased. In particular the search time for the Levy strategy is minimized when $\mu=2$, i--e when the length of the ballistic excursions has a finite second moment so that the walk is no longer of Levy type. As seen in Fig. \ref{fig:levy} the optimal persistent random walk strategy therefore yields a search time shorter  than any Levy walk strategy. 

This optimal persistent search  strategy is in marked contrast with  the simple ballistic motion obtained in  the case of Poisson distributed targets, and shows that the distribution of targets plays a crucial role in the random search problem. In particular, in the biologically relevant cases  of either a single target or   patterns of targets characterized by a peaked distribution of the target to target distance, we find that, as opposed to repeated statements in the literature,  persistent random walks with an exponential distribution of excursion lengths can minimize the search time,  and in that sense perform better than any Levy walk.



\begin{thebibliography}{31}
\expandafter\ifx\csname natexlab\endcsname\relax\def\natexlab#1{#1}\fi
\expandafter\ifx\csname bibnamefont\endcsname\relax
  \def\bibnamefont#1{#1}\fi
\expandafter\ifx\csname bibfnamefont\endcsname\relax
  \def\bibfnamefont#1{#1}\fi
\expandafter\ifx\csname citenamefont\endcsname\relax
  \def\citenamefont#1{#1}\fi
\expandafter\ifx\csname url\endcsname\relax
  \def\url#1{\texttt{#1}}\fi
\expandafter\ifx\csname urlprefix\endcsname\relax\def\urlprefix{URL }\fi
\providecommand{\bibinfo}[2]{#2}
\providecommand{\eprint}[2][]{\url{#2}}

\bibitem[{\citenamefont{B{\'e}nichou et~al.}(2011)\citenamefont{B{\'e}nichou,
  Loverdo, Moreau, and Voituriez}}]{Benichou:2011fk}
\bibinfo{author}{\bibfnamefont{O.}~\bibnamefont{B{\'e}nichou}},
  \bibinfo{author}{\bibfnamefont{C.}~\bibnamefont{Loverdo}},
  \bibinfo{author}{\bibfnamefont{M.}~\bibnamefont{Moreau}}, \bibnamefont{and}
  \bibinfo{author}{\bibfnamefont{R.}~\bibnamefont{Voituriez}},
  \bibinfo{journal}{Reviews of Modern Physics} \textbf{\bibinfo{volume}{83}},
  \bibinfo{pages}{81} (\bibinfo{year}{2011});
  \bibinfo{journal}{Physical Chemistry Chemical Physics}
  \textbf{\bibinfo{volume}{10}}, \bibinfo{pages}{7059} (\bibinfo{year}{2008}).

\bibitem[{\citenamefont{Bell}(1991)}]{Bell:1991}
\bibinfo{author}{\bibfnamefont{J.~W.} \bibnamefont{Bell}},
  \emph{\bibinfo{title}{Searching behaviour, the behavioural ecology of finding
  resources, Animal Behaviour Series}} (\bibinfo{publisher}{Chapman and Hall,
  London}, \bibinfo{year}{1991}).

\bibitem[{\citenamefont{Klafter et~al.}(1996)\citenamefont{Klafter,
  Schlesinger, and Zumofen}}]{Klafter:1996}
\bibinfo{author}{\bibfnamefont{J.}~\bibnamefont{Klafter}},
  \bibinfo{author}{\bibfnamefont{M.}~\bibnamefont{Schlesinger}},
  \bibnamefont{and} \bibinfo{author}{\bibfnamefont{G.}~\bibnamefont{Zumofen}},
  \bibinfo{journal}{Phys. Today} \textbf{\bibinfo{volume}{49}},
  \bibinfo{pages}{33} (\bibinfo{year}{1996}).

\bibitem[{\citenamefont{Viswanathan et~al.}(1996)\citenamefont{Viswanathan,
  Afanasyev, Buldyrev, Murphy, Prince, and Stanley}}]{Viswanathan:1996}
\bibinfo{author}{\bibfnamefont{G.~M.} \bibnamefont{Viswanathan et al.}},
 \bibinfo{journal}{Nature}
  \textbf{\bibinfo{volume}{381}}, \bibinfo{pages}{413} (\bibinfo{year}{1996}).

\bibitem[{\citenamefont{Viswanathan et~al.}(1999)\citenamefont{Viswanathan,
  Buldyrev, Havlin, da~Luz, Raposo, and Stanley}}]{Viswanathan:1999a}
\bibinfo{author}{\bibfnamefont{G.~M.} \bibnamefont{Viswanathan et al.}},
 \bibinfo{journal}{Nature}
  \textbf{\bibinfo{volume}{401}}, \bibinfo{pages}{911} (\bibinfo{year}{1999}).

\bibitem[{\citenamefont{Benichou
  et~al.}(2005{\natexlab{a}})\citenamefont{Benichou, Coppey, Moreau, Suet, and
  Voituriez}}]{Benichou:2005qd}
\bibinfo{author}{\bibfnamefont{O.}~\bibnamefont{Benichou et al.}},
  \bibinfo{journal}{Phys Rev Lett} \textbf{\bibinfo{volume}{94}},
  \bibinfo{pages}{198101} (\bibinfo{year}{2005}{\natexlab{a}});
  \bibinfo{journal}{J. Phys. Condens. Matter}
  \textbf{\bibinfo{volume}{17}}, \bibinfo{pages}{S4275}
  (\bibinfo{year}{2005}{\natexlab{b}});
\bibinfo{author}{\bibfnamefont{O.}~\bibnamefont{Benichou}},
  \bibinfo{author}{\bibfnamefont{C.}~\bibnamefont{Loverdo}},
  \bibinfo{author}{\bibfnamefont{M.}~\bibnamefont{Moreau}}, \bibnamefont{and}
  \bibinfo{author}{\bibfnamefont{R.}~\bibnamefont{Voituriez}},
  \bibinfo{journal}{Phys Rev E }
  \textbf{\bibinfo{volume}{74}}, \bibinfo{pages}{020102}
  (\bibinfo{year}{2006});
  \bibinfo{journal}{J. Phys. Condens. Matter}
  \textbf{\bibinfo{volume}{19}}, \bibinfo{pages}{065141}
  (\bibinfo{year}{2007});
  \bibinfo{author}{\bibfnamefont{C.}~\bibnamefont{Loverdo}},
  \bibinfo{author}{\bibfnamefont{O.}~\bibnamefont{Benichou}},
  \bibinfo{author}{\bibfnamefont{M.}~\bibnamefont{Moreau}}, \bibnamefont{and}
  \bibinfo{author}{\bibfnamefont{R.}~\bibnamefont{Voituriez}},
  \bibinfo{journal}{Phys. Rev. E } \textbf{\bibinfo{volume}{80}}, \bibinfo{pages}{031146}
  (\bibinfo{year}{2009}).

\bibitem[{\citenamefont{Oshanin et~al.}(2007)\citenamefont{Oshanin, Wio,
  Lindenberg, and Burlatsky}}]{Oshanin:2007a}
\bibinfo{author}{\bibfnamefont{G.}~\bibnamefont{Oshanin}},
  \bibinfo{author}{\bibfnamefont{H.~S.} \bibnamefont{Wio}},
  \bibinfo{author}{\bibfnamefont{K.}~\bibnamefont{Lindenberg}},
  \bibnamefont{and} \bibinfo{author}{\bibfnamefont{S.~F.}
  \bibnamefont{Burlatsky}}, \bibinfo{journal}{J. Phys. Condens.
  Matter} \textbf{\bibinfo{volume}{19}}, \bibinfo{pages}{065142}
  (\bibinfo{year}{2007}).

\bibitem[{\citenamefont{Friedrich}(2008)}]{Friedrich:2008kx}
\bibinfo{author}{\bibfnamefont{B.~M.} \bibnamefont{Friedrich}},
  \bibinfo{journal}{Physical Biology} \textbf{\bibinfo{volume}{5}},
  \bibinfo{pages}{026007} (\bibinfo{year}{2008}).

\bibitem[{\citenamefont{Lomholt et~al.}(2008)\citenamefont{Lomholt, Tal,
  Metzler, and Joseph}}]{Lomholt:2008}
\bibinfo{author}{\bibfnamefont{M.~A.} \bibnamefont{Lomholt}},
  \bibinfo{author}{\bibfnamefont{K.}~\bibnamefont{Tal}},
  \bibinfo{author}{\bibfnamefont{R.}~\bibnamefont{Metzler}}, \bibnamefont{and}
  \bibinfo{author}{\bibfnamefont{J.}~\bibnamefont{Klafter}},
  \bibinfo{journal}{PNAS}
  \textbf{\bibinfo{volume}{105}}, \bibinfo{pages}{11055}
  (\bibinfo{year}{2008}).


\bibitem[{\citenamefont{Edwards et~al.}(2007)\citenamefont{Edwards, Phillips,
  Watkins, Freeman, Murphy, Afanasyev, Buldyrev, da~Luz, Raposo, Stanley
  et~al.}}]{Edwards:2007}
\bibinfo{author}{\bibfnamefont{A.~M.} \bibnamefont{Edwards et al.}},
   \bibnamefont{et~al.}, \bibinfo{journal}{Nature}
  \textbf{\bibinfo{volume}{449}}, \bibinfo{pages}{1044} (\bibinfo{year}{2007}).

\bibitem[{\citenamefont{Benhamou}(2007)}]{Benhamou:2007fk}
\bibinfo{author}{\bibfnamefont{S.}~\bibnamefont{Benhamou}},
  \bibinfo{journal}{Ecology} \textbf{\bibinfo{volume}{88}},
  \bibinfo{pages}{1962} (\bibinfo{year}{2007}).

\bibitem[{\citenamefont{James et~al.}(2011)\citenamefont{James, Plank, and
  Edwards}}]{James:2011uq}
\bibinfo{author}{\bibfnamefont{A.}~\bibnamefont{James}},
  \bibinfo{author}{\bibfnamefont{M.~J.} \bibnamefont{Plank}}, \bibnamefont{and}
  \bibinfo{author}{\bibfnamefont{A.~M.} \bibnamefont{Edwards}},
  \bibinfo{journal}{Journal of The Royal Society Interface}
  \textbf{\bibinfo{volume}{8}}, \bibinfo{pages}{1233} (\bibinfo{year}{2011}).

\bibitem[{\citenamefont{Redner}(2001)}]{Redner:2001a}
\bibinfo{author}{\bibfnamefont{S.}~\bibnamefont{Redner}},
  \emph{\bibinfo{title}{A guide to first passage time processes}}
  (\bibinfo{publisher}{Cambridge University Press, Cambridge, England},
  \bibinfo{year}{2001}).

\bibitem[{\citenamefont{Condamin et~al.}(2007)\citenamefont{Condamin, Benichou,
  Tejedor, Voituriez, and Klafter}}]{Condamin:2007zl}
\bibinfo{author}{\bibfnamefont{S.}~\bibnamefont{Condamin et al.}},
  \bibinfo{journal}{Nature} \textbf{\bibinfo{volume}{450}}, \bibinfo{pages}{77}
  (\bibinfo{year}{2007}).

\bibitem[{\citenamefont{B{\'e}nichou et~al.}(2010)\citenamefont{B{\'e}nichou,
  Chevalier, Klafter, Meyer, and Voituriez}}]{BenichouO.:2010}
\bibinfo{author}{\bibfnamefont{O.}~\bibnamefont{B{\'e}nichou et al.}},
  \bibinfo{journal}{Nat Chem} \textbf{\bibinfo{volume}{2}},
  \bibinfo{pages}{472} (\bibinfo{year}{2010}).

\bibitem[{\citenamefont{Berg}(2004)}]{bergcoli}
\bibinfo{author}{\bibfnamefont{H.}~\bibnamefont{Berg}},
  \emph{\bibinfo{title}{E. Coli in Motion}} (\bibinfo{publisher}{Springer, New
  York}, \bibinfo{year}{2004}).

\bibitem[{\citenamefont{Blanco and Fournier}(2003)}]{Blanco:2003a}
\bibinfo{author}{\bibfnamefont{S.}~\bibnamefont{Blanco}} \bibnamefont{and}
  \bibinfo{author}{\bibfnamefont{R.}~\bibnamefont{Fournier}},
  \bibinfo{journal}{EPL } \textbf{\bibinfo{volume}{61}},
  \bibinfo{pages}{168} (\bibinfo{year}{2003}).

\bibitem[{\citenamefont{Mazzolo}(2004)}]{Mazzolo:2004a}
\bibinfo{author}{\bibfnamefont{A.}~\bibnamefont{Mazzolo}},
  \bibinfo{journal}{EPL } \textbf{\bibinfo{volume}{68}},
  \bibinfo{pages}{350} (\bibinfo{year}{2004}).

\bibitem[{\citenamefont{Zoia et~al.}(2011)\citenamefont{Zoia, Dumonteil, and
  Mazzolo}}]{Zoia:2011vn}
\bibinfo{author}{\bibfnamefont{A.}~\bibnamefont{Zoia}},
  \bibinfo{author}{\bibfnamefont{E.}~\bibnamefont{Dumonteil}},
  \bibnamefont{and} \bibinfo{author}{\bibfnamefont{A.}~\bibnamefont{Mazzolo}},
  \bibinfo{journal}{Phys Rev Lett} \textbf{\bibinfo{volume}{106}},
  \bibinfo{pages}{220602} (\bibinfo{year}{2011}).

\bibitem[{\citenamefont{Ernst}(1988)}]{Ernst:1988fk}
\bibinfo{author}{\bibfnamefont{M.~H.} \bibnamefont{Ernst}},
  \bibinfo{journal}{J. Stat. Phys.}
  \textbf{\bibinfo{volume}{53}}, \bibinfo{pages}{191} (\bibinfo{year}{1988}).

\bibitem[{\citenamefont{Gilbert et~al.}(2011)\citenamefont{Gilbert, Nguyen, and
  Sanders}}]{Gilbert:2011ys}
\bibinfo{author}{\bibfnamefont{T.}~\bibnamefont{Gilbert}},
  \bibinfo{author}{\bibfnamefont{H.~C.} \bibnamefont{Nguyen}},
  \bibnamefont{and} \bibinfo{author}{\bibfnamefont{D.~P.}
  \bibnamefont{Sanders}}, \bibinfo{journal}{J.  Phys. A} \textbf{\bibinfo{volume}{44}} (\bibinfo{year}{2011}).

\bibitem[{\citenamefont{B\'enichou et~al.}(2005)\citenamefont{B\'enichou,
  Coppey, Moreau, Suet, and Voituriez}}]{Benichou:2005a}
\bibinfo{author}{\bibfnamefont{O.}~\bibnamefont{B\'enichou et al.}},
  \bibinfo{journal}{EPL } \textbf{\bibinfo{volume}{70}},
  \bibinfo{pages}{42} (\bibinfo{year}{2005}).

\bibitem[{\citenamefont{Condamin et~al.}(2005)\citenamefont{Condamin, Benichou,
  and Moreau}}]{Condamin:2005qr}
\bibinfo{author}{\bibfnamefont{S.}~\bibnamefont{Condamin}},
  \bibinfo{author}{\bibfnamefont{O.}~\bibnamefont{Benichou}}, \bibnamefont{and}
  \bibinfo{author}{\bibfnamefont{M.}~\bibnamefont{Moreau}},
  \bibinfo{journal}{Phys Rev E }
  \textbf{\bibinfo{volume}{72}}, \bibinfo{pages}{016127}
  (\bibinfo{year}{2005}).

\bibitem[{\citenamefont{Aldous and Fill}(1999)}]{Aldous:1999}
\bibinfo{author}{\bibfnamefont{D.}~\bibnamefont{Aldous}} \bibnamefont{and}
  \bibinfo{author}{\bibfnamefont{J.}~\bibnamefont{Fill}},
  \emph{\bibinfo{title}{Reversible Markov chains and random walks on graphs}}
  (\bibinfo{publisher}{http://www.stat.berkeley.edu/users/aldous/%
}, \bibinfo{year}{1999}).


\bibitem[{\citenamefont{Condamin and Benichou}(2006)}]{Condamin:2006fp}
\bibinfo{author}{\bibfnamefont{S.}~\bibnamefont{Condamin}} \bibnamefont{and}
  \bibinfo{author}{\bibfnamefont{O.}~\bibnamefont{Benichou}},
  \bibinfo{journal}{J Chem Phys} \textbf{\bibinfo{volume}{124}},
  \bibinfo{pages}{206103} (\bibinfo{year}{2006}).

\bibitem[{\citenamefont{Tejedor et~al.}(2011)\citenamefont{Tejedor,
  B{\'e}nichou, Metzler, and Voituriez}}]{Tejedor:2011uq}
\bibinfo{author}{\bibfnamefont{V.}~\bibnamefont{Tejedor}},
  \bibinfo{author}{\bibfnamefont{O.}~\bibnamefont{B{\'e}nichou}},
  \bibinfo{author}{\bibfnamefont{R.}~\bibnamefont{Metzler}}, \bibnamefont{and}
  \bibinfo{author}{\bibfnamefont{R.}~\bibnamefont{Voituriez}},
  \bibinfo{journal}{J.  Phys. A}
  \textbf{\bibinfo{volume}{44}}, \bibinfo{pages}{255003}
  (\bibinfo{year}{2011}).

\bibitem[{\citenamefont{Tejedor et~al.}(2009)\citenamefont{Tejedor,
  B{\'e}nichou, and Voituriez}}]{Tejedor:2009}
\bibinfo{author}{\bibfnamefont{V.}~\bibnamefont{Tejedor}},
  \bibinfo{author}{\bibfnamefont{O.}~\bibnamefont{B{\'e}nichou}},
  \bibnamefont{and}
  \bibinfo{author}{\bibfnamefont{R.}~\bibnamefont{Voituriez}},
  \bibinfo{journal}{Phys Rev E} \textbf{\bibinfo{volume}{80}}
  (\bibinfo{year}{2009}).

\end{thebibliography}

\end{document}